\documentclass[12pt]{article}
\usepackage[francais,german,english]{babel}
\usepackage{amssymb}
\def \ub {\underline}

\begin{document}

\thispagestyle{empty}

\baselineskip=.75truecm

\selectlanguage{english}

\rightline{{\bf ETH-TH/96-07\ }}

\rightline{March 1996\qquad\quad}

\vspace{2cm}

\begin{center}
{\Large{\bf Multilinear Evolution Equations\\
for Time-Harmonic Flows\\
\smallskip
in Conformally Flat Manifolds}}
\end{center}

\vspace{1.5cm}

\begin{center}
{\large{\bf Jens
    Hoppe}}\renewcommand{\thefootnote}{\fnsymbol{footnote}}\footnote[1]{\ 
Heisenberg Fellow\\  
On leave of absence from Karlsruhe
University}\renewcommand{\thefootnote}{\arabic{footnote}}\\ 
{\large{\bf Institut f\"ur Theoretische Physik\\
ETH-H\"onggerberg\\
CH-8093 \ Z\"urich}}
\end{center}

\vspace{5cm}

\noindent {\bf Abstract} \ \ It is shown that time-harmonic
hypersurface motions in various, conformally flat, $N$-dimensional
manifolds admit a multilinear description, \ $\dot{L} = \{ L,
M_1, \cdots, M_{N-2}\}$, automatically generating infinitely many
conserved quantities, as well as leading to new (integrable) matrix
equations. Interestingly, the conformal factor can be changed without
changing $L$.

\vfill\eject

\noindent For hypersurface motions in $\mathbb{R}^N$, having the property
that the time at which a point in space is reached is a harmonic
function, a multilinear description, \ $
\dot{L}\;=\;\{ L, M_1, M_2, \cdots,$\break
$ M_{N-2} \} \ $, was given in
[1], with $L$ and the $M$'s being linear in the embedding functions \
$x^i (t, \varphi^1,\cdots,\varphi^{N-1})$ \ (the coordinates of the
hypersurface $\Sigma_t$ in $\mathbb{R}^N$). \ In this letter, it is
pointed out that by non-linearly deforming $M_1,\cdots, M_{N-2}$ \
(while keeping $L$ fixed!) one may effectively deform the embedding
space into a curved one.

To start with a more technical (but needed) result, let me
consider hypersurface motions (in a Riemannian manifold, $\cal{N}$) of
the form 
\begin{equation}
\dot x^i\;=\;\sqrt{g}/ \rho\;\left( h(x)n^i\,+\,f^{ij}(x)\;\zeta_{jk}
  (x) n^k\right)\quad i\;=\; 1,2,\cdots, N \ ,
\end{equation}
where $n^i$ denotes the hypersurface-normal (of given
orientation), \ $x^i (t,\varphi^1,\cdots,\varphi^M)$ \ the (closed)
hypersurface $\Sigma_t$ (in a parametric description),\
$\varphi^1,\cdots,\varphi^M$ \ being local coordinates on a (compact)
$M=N-1$ dimensional Riemannian manifold $\Sigma$ (on which the $x^i$
are timedependent functions), $\sqrt{g}$ \ the volume density (on
$\Sigma_t$) induced by the embedding Riemannian manifold $\cal{N}$
(with metric $\zeta_{ij} (x))$,
\begin{equation}
g\;=\;\det \ \left( \zeta_{ij} (x) \ \frac{\partial x^i}{\partial
    \varphi^r} \ \frac{\partial x^j}{\partial
    \varphi^s}\right)_{r,s\,=\,1\cdots M} \ ,
\end{equation}
$\rho=\rho (\varphi^1\cdots\varphi^M$) \ some time-independent
(positive) density on $\Sigma$ (reflecting the topology, and making
$\sqrt{g\;}/\rho$ a well defined function, rather than a density, on
$\Sigma$~\renewcommand{\thefootnote}{\fnsymbol{footnote}}\footnote[1]{
  but {\bf not} on $\Sigma_t$; so, in contrast with 
  commonly considered flows, (1) requires as initial conditions
  $\Sigma_{t=0}$ \ {\bf and} an initial velocity field consistent with
  (1) (or, equivalently, a given {\bf parametrized} initial
  hypersurface).}\renewcommand{\thefootnote}{\arabic{footnote}} and,
finally, $h(x)$ and $f^{ij}(x)=\,-\,f^{ji}(x)$ 
some function, resp. antisymmetric tensor, on $\cal{N}$ satisfying
\begin{equation}
\nabla_j \left( \zeta^{ij}\,h\;+\;f^{ij}\right)\;=\;0 \ .
\end{equation}
(3) implies the existence of infinitely many conserved quantities for
the flow (1), as can easily be seen as follows: \ Let $Q$ be a
harmonic function (of $x$) inside some given initial hypersurface \
$\Sigma_{t=0}$; \ then
\begin{eqnarray}
&&\frac{\partial}{\partial t}\ \int d^M \varphi \ \rho (\varphi)\ Q\
\left( x (t,\varphi)\right) \nonumber \\ 
&& = \ \int d^M\varphi \ \frac{\partial Q}{\partial x^i} \ \left(
  \zeta^{ij} h\,+\, f^{ij}\right)\ n_j\ \sqrt{g} \\
&& = \ \int d^N\,x\,\sqrt{\zeta} \ \nabla_j\ \left( \nabla_i Q\
  (\zeta^{ij} h\,+\,f^{ij})\right)\;=\; 0 \ ,\nonumber
\end{eqnarray}
using $\nabla^i\ \nabla_i\ Q = 0$, the antisymmetry of \ $f^{ij}$, and
(the normal component of) (3). As the $f^{ij}$-part of (1) is purely
tangential (hence, in principle, can be gotten rid of by
reparametrizing the surface), and having shown in [2] that purely
normal motions with covariantly constant $h(x)$ have integrals of
harmonic functions as conserved quantities, a naive guess would be
that the motions (1) are geometrically equivalent to those with
\begin{equation}
h(x)\;\equiv\;{\rm const.} \quad , \quad f^{ij} (x) \;\equiv\; 0
\end{equation}
(trivially satisfying (3)). In fact, I was led to the motions (1) when
considering Lax-tuples for the case (5), with $\zeta_{ij} (x) =
\delta_{ij}$.

However, a more interesting situation prevails, as is shown by the
following theorem:

\medskip

Let $\tilde{\varphi}^1,\cdots,\tilde{\varphi}^M$ \ be local parameters
on $\Sigma$, and
\begin{equation}
\frac{\partial x^i}{\partial \tilde{t}}\;=\; h^2\,(x)\;
\frac{\sqrt{\tilde{g}}}{\tilde{\rho}}\; n^i
\end{equation}
define a hypersurface motion in ($\cal{N}$, $\zeta_{ij})$, where \
$\tilde{\rho} (\tilde{\varphi}^1,\cdots,\tilde{\varphi}^M)$ \ is some
time-independent density on $\Sigma$. Then (provided (3) holds) there
exists a time-dependent reparametrisation of
$\widetilde{\Sigma}_{\tilde{t}}$ \ (and a constant rescaling of the
time),
\begin{equation}
\tilde{t},\tilde{\varphi}^1,\cdots,\tilde{\varphi}^M\, \to\,
t\;=\;\lambda\tilde{t}, \ \varphi^r\,=\,\varphi^r(\tilde{t},
\tilde{\varphi}^1,\cdots,\tilde{\varphi}^M) \ (r\,=\,1,\cdots, M) 
\end{equation}
such that the $x^i$, expressed as functions of the `new' parameters,
$(t,\varphi)$, satisfy (1).
More explicitely: Consider a particular solution of (6), and a
reparametrisation of $\widetilde{\Sigma}_{\tilde{t}\,=\,0}$ satisfying
\begin{equation}
\rho(\varphi)\,{\rm det \ }\left( \frac{\partial \varphi^r}{\partial
    \tilde{\varphi}^s}\right) \ = \
  \frac{\lambda\,\cdot\,\tilde{\rho}\,(\tilde{\varphi})}{h\,\left(
      x\,(\tilde{t},\tilde{\varphi})\right)} 
\end{equation}
at time $\tilde{t}=0$ (here, the freedom of choosing $\lambda$ is
needed).

\medskip

Let the time-evolution of the $\varphi^r$ be given by the equation 
\begin{equation}
\partial_{\tilde{t}}\,\varphi^r\;=\;
\frac{1}{(M-1)!}\,
\ f_{i_2\cdots i_M}(x) \ h(x)\ \frac{1}{\tilde{\rho}
  (\tilde{\varphi})} \ 
\bigl\{ \varphi^r, x^{i_2}, \cdots ,
  x^{i_M}\mathop{\big\}}\limits_{\widetilde{ }}  \ ,
\end{equation}
where \ $x\,=\,x (\tilde{t},\tilde{\varphi})$,
\begin{equation}
f^{ij}\;=:\; \frac{\epsilon^{ij\;i_2\cdots i_M}}{\sqrt{\zeta}\,(M-1)!}
\ f_{i_2 \cdots i_M} \ , 
\end{equation}
and
\begin{equation}
\bigl\{ f_1, \cdots , f_M\mathop{\big\}}\limits_{\widetilde{ }} \;:=\;
\epsilon^{r_1\cdots r_M} \ 
\frac{\partial f_1}{\partial \tilde{\varphi}^1} \ \cdots \
\frac{\partial f_M}{\partial \tilde{\varphi}^M} 
\end{equation}
(for any set of $M$ functions on $\Sigma$).

Then the solutions of (9), \ $\varphi^r (\tilde{t},\tilde{\varphi})$,
satisfying (8) at time $\tilde{t}=0$, will satisfy (8) also  for
$\tilde{t}>0$, and $x^i (\tilde{t},\tilde{\varphi})$, when expressed as
functions of $t$ and $\varphi^r \ (r=1,2,\cdots,M)$, will solve
(1). 

Hence, geometrically, (1) and (6) are (up to the constant
rescaling of $t$) equivalent. 

In particular, 
\begin{equation}
\nabla^i \ \left( h^2 \ (x) \ \nabla_i\ t(x)\right)\;=\; 0
\end{equation}
when $t$ is expressed as a function of the $x^i$ by considering the
transformation 
\begin{equation}
t, \varphi^1, \cdots, \varphi^M\ \to \ x^i\,=\,x^i\;(t, \varphi^1,
\cdots, \varphi^M)
\end{equation}
for equation (1).

As for the proof, one first shows that a reparametrisation of the form
(7) will transform (6) into (1), provided (8) and (9) hold: use \
$\partial_{\tilde{t}} = \lambda \;\partial_t\,+\,(\partial_{\tilde{t}}
\varphi^r)\ \partial_r$ \ and
\begin{equation}
\sqrt{\tilde{g}\;}/\tilde{\rho} \ = \ \sqrt{g\;}/\rho \ \frac{\lambda}{h\,(x)}
\end{equation}
to get the condition
\begin{eqnarray}
\partial_{\tilde{t}} \varphi^r\;\partial_r x^i &=&
\frac{\lambda}{(M-1)!} \ f_{i_2 \cdots i_M} \ \frac 1 \rho \
\left\{ x^i, x^{i_2}, \cdots , x^{i_M} \right\} \nonumber \\
&=& \frac{1}{(M-1)!} \ f_{i_2 \cdots i_M} \ h \
\frac{1}{\tilde{\rho}} \ \bigl\{ x^i, x^{i_2}, \cdots, x^{i_M}
\mathop{\big\}}\limits_{\widetilde{ }} \ ;
\end{eqnarray}
comparing the coefficients of $\tilde{\partial}_s x^i$ \ on both sides
(using  \ $\partial_r x^i\,=\,\left(\left(
    \frac{\partial\varphi}{\partial\tilde{\varphi}}\right)^{-1}\right)_r^s \   
\tilde{\partial}_s x^i$),  
\ and multiplying by \ $\frac{\partial \varphi^{r'}}{\partial\tilde{\varphi}^s}$, 
\ one gets (9). The crucial step then is to show the 
consistency of (9) with (8) (provided (3) holds) 
by calculating the time-derivative of (8), using (9):

On the one hand, 
\begin{equation}
\partial_{\tilde{t}} \ \left( \frac{\lambda}{\rho (\varphi) h
    (x)}\right) \;=\; \lambda\ \sqrt{\tilde{g}\;}/\tilde{\rho} \ h^2
(x)\ n^i \ \nabla_i \ \left( \frac{1}{\rho \,h}\right) \ ,
\end{equation}
using the transformation
\begin{equation}
\tilde{t}, \ \tilde{\varphi}^1, \cdots, \tilde{\varphi}^M \ \to \
x^i\;=\; x^i (\tilde{t}, \tilde{\varphi})
\end{equation}
and (6); on the other hand, using (9), one finds
\begin{eqnarray}
&&\partial_{\tilde{t}} \ \left( \frac{1}{\tilde{\rho}
    (\tilde{\varphi})} \ \bigl\{ \varphi^1, \cdots,
  \varphi^M\mathop{\big\}}\limits_{\widetilde{ }} 
\right) \nonumber \\
&& = \ \frac{1}{\left( (M-1)!\right)^2} \ \epsilon_{r_1\cdots r_M} \
\left( \begin{array}{c}
\frac{1}{\tilde\rho}\bigl\{ \varphi^{r_1}, x^{i_2}, \cdots, x^{i_M}
  \mathop{\big\}}\limits_{\widetilde{ }} \bigl\{ \sqrt{\zeta}
  \,h\,f_{i_2\cdots i_M}, 
    \varphi^{r_2},\cdots
    \varphi^{r_M}\mathop{\big\}}\limits_{\widetilde{ }} \\
+\; \frac{1}{\tilde\rho}\ \biggl\{ \frac{1}{\tilde\rho} \bigl\{
    \varphi^{r_1}, x^{i_2},\cdots,
    x^{i_M}\mathop{\big\}}\limits_{\widetilde{ }} , \
  \varphi^{r_2},\cdots,
  \varphi^{r_M}\mathop{\bigg\}}\limits_{\widetilde{ }}\; \cdot \
  \sqrt{\zeta} \ f_{i_2 \cdots i_M} \ h  
\end{array} \right) \nonumber \\
&&=\ \frac{\lambda}{(M-1)!} \ \frac{1}{\tilde\rho} \ \bigl\{
  \sqrt{\zeta}\;h\;f_{i_2 \cdots i_M}, x^{i_2}, \cdots, x^{i_M}
\mathop{\big\}}\limits_{\widetilde{ }} \; \cdot \;
\frac{1}{\rho(\varphi) h (x)} \\ 
&&+\ \frac{\lambda}{(M-1)!} \ \frac{1}{\tilde\rho} \ \biggl\{
  \frac{1}{\rho(\varphi)h(x)}, x^{i_2}, \cdots ,
  x^{i_M}\mathop{\bigg\}}\limits_{\widetilde{ }} \ \cdot\ \sqrt{\zeta} \
  f_{i_2\cdots i_M} h \nonumber
\end{eqnarray}
where for the second equality it is easiest to use
\begin{equation}
\frac{1}{\tilde\rho}\ \bigl\{ \qquad
\mathop{\big\}}\limits_{\widetilde{ }} \ = \ \frac \lambda h \ 
\frac 1 \rho \ \left\{ \ \cdots \ \right\}
\end{equation}
intermediately; the terms proportional to \ $\left\{ h,
  x^{i_2},\cdots, x^{i_M}\right\}$ \ in (18) cancel, and the equality
of (18) and (17) easily follows when using (3),
\begin{eqnarray}
&&\nabla_i f^{ij}\;=\;\frac{1}{\sqrt{\zeta}} \ \partial_i \left(
  \sqrt{\zeta}\ f^{ij}\right) \left( f^{ij}\;=\;-\;f^{ji}\right)\ ,
\nonumber \\
&&\bigl\{ x^{i_1}, x^{i_2},\cdots,
x^{i_M}\mathop{\big\}}\limits_{\widetilde{ }} \
\;=\;\frac{\epsilon^{j\,i_1\cdots i_M}}{\sqrt{\zeta}} \ n_j\ \sqrt{\tilde{g}}
\end{eqnarray}
 and (for the terms
containing derivatives of \ $\frac{1}{\rho(\varphi)}$)
\begin{equation}
\left( h (x) n^i\;+\;f_j^i (x) n^j\right) \ \nabla_i\;\varphi^r\ = \ 0
\end{equation}
(which is simply \ $\partial_t \varphi^r = 0$ in (1), using (13)).

For (12), one can either use the results of [2], applied to (6)
(i.e. already using that (1) and (6) are related via (7), when (3)
holds), -- or, using $n_i=\partial_it \left(\zeta^{jk}
  \partial_jt\,\partial_k\,t\right)^{-\,1/2}$ \ and (from (1))
\begin{equation}
\sqrt{g} / \rho \;=\; \frac{1}{h (x)} \ \left(
  \dot{x}^i\;n_i\right)\;=\; \frac{1}{h (x) |\nabla t|} \ ,
\end{equation}
derive from (1) an equation containing only $f^{ij}, h, \partial_it,
\zeta_{ij}$ \ (and their $x$-derivatives), using the $x^i$ (cp. (13))
as independent variables (cp. [2] for $f^{ij} \equiv 0$). The latter
procedure provides an independent check on the crucial relative sign
of $h$ and $f^{ij}$ \ (for \ $\nabla_j \left( \zeta^{ij} h -
  f^{ij}\right)=0$ \ one could multiply the r.h.s. of (8), and the
l.h.s. of (9), by $h^2(x)$, to obtain $\dot{x}^i =
\frac{\sqrt{\tilde{g}}}{\tilde{\rho}} \ n^i$, resp. $\nabla_i \nabla^i
t$ from (1), via (7)),
but is quite cumbersome, and shall only be sketched for $N=3$,
$\zeta_{ij}=\delta_{ij}$, and differentiating solely the normal
component of (1), resp.
\begin{equation}
\stackrel{\cdot}{\vec{x}}\;=\;\frac{1}{|\nabla t|} \left( \vec{n} +
  \vec{n} \times \vec{f}/h\right) \ ,
\end{equation}
with respect to $t$. Using $\partial_t =
\stackrel{\cdot}{\vec{x}}\,\vec{\nabla}$ one obtains 
\begin{equation}
\frac{1}{|\nabla \,t|} \ \left( \vec{n} +
  \vec{n}\times\vec{f}/h\right)\ \cdot \ \vec{\nabla} \ \left(
  \frac{1}{|\nabla\,t|}\right) \ ,
\end{equation}
while
\begin{eqnarray}
&&\partial_t \left( h \sqrt{g}/\rho\right) = \left(
  \stackrel{\cdot}{\vec{x}} \vec{\nabla} h\right) \ \sqrt{g}/\rho + h
\ \dot{\sqrt{g}}/\rho \nonumber\\
&& = \ \frac{1}{|\nabla\,t|} \ \left( \vec{n} + \vec{n}
  \times\vec{f}/h\right)\ \cdot \ \vec{\nabla}\,h \ \cdot \ \frac{1}{h
  |\nabla\,t|} \nonumber\\
&&+ \ \frac{1}{|\nabla\,t|^2} \ \vec{\nabla} \ \left(
  \frac{\vec{\nabla}\,t}{|\nabla\,t|} \right) \\
&&- \ \frac{1}{|\nabla\,t|} \ \vec{f}\ \cdot \ \left(
  \vec{n}\times\vec{\nabla}\left( \frac{1}{h|\nabla\,t|}\right)\right)
\nonumber\\
&& - \ \frac{1}{|\nabla\,t|^2\;h}\ \vec{n} \ \cdot \ \left(
  \vec{\nabla} \times \vec{f}\right) \ . \nonumber 
\end{eqnarray} 
Comparing (24) and (25) one obtains
\begin{equation}
\vec{\nabla} \left( h (\vec{x}) \vec{\nabla}\,t\right) \ - \ \left(
  \vec{\nabla} \times \vec{f}\right) \ \cdot \ \vec{\nabla}\,t \
\stackrel{!}{=} \ 0 
\end{equation}
which indeed gives (12), when using (3) (i.e. $\vec{\nabla} h +
\vec{\nabla} \times \vec{f} = 0$).

For the derivation of (25), it is perhaps useful to note the general
formula
\begin{equation}
\frac{\partial}{\partial\,t} \left( \ln \sqrt{g}\right)\ = \ \left(
  \delta_j^i - n^i \, n_j\right) \ \nabla_i \dot{x}^j
\end{equation}
(in the case of purely normal hypersurface motion, $\dot{x}^j = v
\cdot n^j$, simply giving $v$ times the mean curvature, $H$, of
$\Sigma_t$) or alternatively (for (1), and in $\mathbb{R}^3$)
\begin{eqnarray}
&& \dot{\sqrt{g}} \ = \ \sqrt{g} \left( \sqrt{g}\frac h \rho \right)\;H
\ - \ \left\{ \vec{f} \sqrt{g}/\rho, \vec{x}\right\} \nonumber\\
&& = \ \sqrt{g} \ \frac{1}{|\nabla\,t|} \ \vec{\nabla}
\;\vec{n}\;-\;\sqrt{g}\;\left( \vec{n} \times \vec{\nabla}\right)
\cdot \left( \vec{f} \frac{\sqrt{g}}{\rho}\right) \ .
\end{eqnarray}
Let me now turn to the discussion of multilinear evolution equations
\begin{equation}
\dot{L} \ = \ \left\{ L, M_1, M_2,\cdots, M_{N-1} \right\} 
\end{equation}
related to (1).

As shown in [1], the equations of motion for a time-harmonic
hypersurface in $\mathbb{R}^N$ [2],
\begin{equation}
\dot{x}_i\;=\;\frac{\epsilon_{ii_1i_2...i_M}}{M!}\ \left\{ x_{i_1},
  x_{i_2},\cdots, x_{i_M}\right\} \ \ \ i\;=\;1,2,\cdots, N \ ,
\end{equation}
(where from now on, the factor $\frac{1}{\rho}$ is included in the
definition of $\{ , \cdots, \}$, and no distinction is made between
upper and lower indices) can be cast into the form (29), automatically
implying the time-independence of 
\begin{equation}
\int \ d^M\varphi \ \rho(\varphi) \ L^n \ ,
\end{equation}
where L (depending on several spectral parameters, ${\rm \lambda_a}$ )
will be of the form [1]
\begin{equation}
L\;=\;\sum_{i=1}^N \ \mathbb{L}_i\;x_i \ , \ \sum_{i=1}^N \
\mathbb{L}_i^2 \;=\; 0 \quad (\mathbb{L}_i\;\in \mathbb{C}) \ .
\end{equation}
For odd $N(\equiv 2m+1)$, e.g., one may choose 
\begin{equation}
L\;=\;\sum_{a=1}^m \ \left( \lambda_a\,z_a\;-\;
  \frac{\overline{z_a}}{\lambda_a}\right) \;+\; 2 \sqrt{m} \ x_N
\end{equation}
(with $z_1 := x_1 + i x_2, \ z_2 := x_3 + i x_4, \cdots$ )
\ while the $M_\alpha$'s \
$(\alpha = 1, \cdots, N-2)$ are linear combinations of $\lambda_a
z_a$, $\frac{\overline{z_a}}{\lambda_a}$ \ , and $x_N$ (i.e. elements
$\vec{M}_\alpha$ of some vector space) satisfying
\begin{equation}
\det \left( \vec{L}\,\vec{M}_1 \cdots \vec{M}_{N-2}\;\vec{e}_j\right)
\;=\; -\,2 \left( \frac i 2 \right)^m \ \hat{\vec{L}} \cdot \vec{e}_j
\end{equation}
where
\begin{equation}
\hat{\vec{L}} \;:=\; -\, \left( L_2, L_1, \cdots, L_{N-1}, L_{N-2}, \
  \frac 1 2 \ L_N \right)
\end{equation}
($ = - (-1, +1, \cdots, -1, +1, \sqrt{m})$ \ for the particular choice
(33); in general, any $\vec{L}$ satisfying $\hat{\vec{L}} \cdot
\vec{L} = 0$ will do, cp. (32)). Suppose now replacing the $z_a$ in
the expressions for $M_1, \cdots, M_{N-2}$ by arbitrary holomorphic
functions $f_a (z_a)$ \ satisfying \ $\overline{f_a (z_a)} = f_a
(\overline{z_a})$, 
and $x_N$ by \ $u(x_1, \cdots x_N)$, while keeping $L$ fixed!

The equations of motion generated via (29) will still be of the form
\begin{equation}
\dot{x}_i \;=\; \eta_{ij}(x_1, \cdots, x_N) \
\frac{\epsilon_{ji_1\cdots i_M}}{M!} \ \left\{
  x_{i_1},\,x_{i_2},\cdots, x_{i_M}\right\} \ i\;=\;1,\cdots, N
\end{equation}
$\left( = \eta_{ij} (x)\;\sqrt{g}/\rho\;n_j\right)$. For the simplest
case, $m=1$, and $(z := x_1 + ix_2)$
\begin{equation}
M\;=\;c\,\cdot\,i\;\left( \lambda\;z\;+\;x_3\right)\;+\; \lambda\,f
(z)\;-\; \frac{\overline{f(z)}}{\lambda} \;+\; 2 u (\vec{x})
\end{equation}
(having separated explicitely the piece leading to \
$\dot{\vec{x}}\;=\;c\,\sqrt{g}/\rho\,\vec{n}$) \ one obtains
\begin{equation}
\eta_{ij}\left( x_1\,x_2\,x_3\right)\;=\;2 \
\left( \begin{array}{c}
\frac{c}{2}\ -\ \partial_2\alpha \qquad
\partial_1\alpha-\partial_3u \quad \partial_2u\\
\partial_3u-\partial_1\alpha \quad
\quad \frac{c}{2}\;-\;\partial_2\alpha \quad -\;\partial_1u \\
- -\;\partial_2 u \qquad \quad\partial_1 u \qquad
\quad \frac{c}{2}\;-\;\partial_2\alpha 
\end{array} \right)\ ,
\end{equation}
where \ $\alpha = \alpha (x_1x_2)$ = ${\cal{R}}e$ $\left( f(z)\right)$; \
hence
\begin{eqnarray}
\dot{\vec{x}}&=& \left(
  c-2\partial_2\alpha\right)\;\sqrt{g}/\rho\,\vec{n}\;+\; 2\;\left\{
  \vec{x}, u(\vec{x})\right\}\;+\;2(\partial_1\alpha)
\{x_3,\vec{x}\}\nonumber \\
&=&\left(
  c-2\partial_2\alpha\right)\;\sqrt{g}/\rho\;\vec{n}\;-\;2\;\sqrt{g}/\rho 
\vec{n}\;\times\; \vec{\nabla}u\;+\;2\;\sqrt{g}/\rho\;\vec{n}\;\times\; 
\left( \begin{array}{c} 
0 \\ 0 \\ \partial_1\alpha
\end{array} \right)\ .
\end{eqnarray}
One should note that via approximating functions on $S^2$ or $T^2$ by
matrices (see e.g. [3]) (37), with $L=\lambda
z\,-\,\frac{\overline{z}}{\lambda} + 2x_3$ (cp. (33)), carries over to
matrix-equations which will be deformations of the `Nahm-equations'
(arising in the context of self-dual Yang-Mills theories and the
construction of monopoles; see e.g. [4]): 

\noindent With
\begin{eqnarray}
L &=& \lambda (X_1+iX_2)\;-\;\frac{(X_1-iX_2)}{\lambda}\;+\;2X_3
\nonumber\\
M &=& c\;\left( \lambda (X_1\;+\;iX_2)\;+\;X_3\right) \\
&& +\; \frac{a\lambda}{2} \;\left( X_1^2\;-\;X_2^2\;+\;i \left(
    X_1X_2\;+\;X_2X_1\right)\right)\nonumber\\
&& -\;\frac{a}{2\lambda}\;\left( X_1^2\;-\;X_2^2\;-\;i\;\left( X_1
    X_2\;+\;X_2X_1\right)\right)\ , \nonumber 
\end{eqnarray}
e.g., \ $\dot{L} = i [L,M]$ \ will be equivalent to the (matrix-)
equations of motion 
\begin{eqnarray}
\dot{X}_1 &=& c\left[ X_2,X_3\right]\;+\;ia \left(X_1\left[
    X_3,X_1\right]\;+\;\left[ X_3,X_1\right] X_1\right)\nonumber \\
&&\qquad\qquad +\;ia\left( X_2\left[ X_2,X_3\right]\;+\;\left[
    X_2,X_3\right] X_2\right)\nonumber \\
\dot{X}_2 &=& c\left[ X_3,X_1\right]\;+\;ia \left(X_2\left[
    X_3,X_1\right]\;+\;\left[ X_3,X_1\right] X_2\right.\nonumber \\
&& \qquad\qquad\left.-\;X_1\left[
    X_2,X_3\right]\;-\;\left[ X_2,X_3\right] X_1\right)\nonumber \\
\dot{X}_3 &=& c\left[ X_1, X_2\right] \\
&&\qquad\qquad +\;ia \left( X_2 \left[ X_1,X_2\right]\;+\;\left[
    X_1,X_2\right] X_2\right) \ , \nonumber
\end{eqnarray}
having $Tr L^n$, i.e. the trace of all symmetric harmonic polynomials
in the 3 matrices \ $X_1(t), X_2(t), X_3(t)$, as conserved quantities -
which for the lowest
order non-trivial ones, like
\begin{equation}
\frac{1}{2} \ Tr L^2|_{\lambda=0}\;=\; Tr \left( 2
  X_3^2\;-\;X_1^2\;-\;X_2^2\right) 
\end{equation}
may easily be seen directly from (41).

Finally, note that (39) is of the form (1), with $N=3$, $\zeta_{ij} =
\delta_{ij}$, $f^{ij} = \epsilon^{ijk} f_k$, and
\begin{eqnarray}
h(x) &=& \left( c \;-\;2 \partial_2 \alpha\right) \nonumber\\
\vec{f}(x) &=& -2\; \vec{\nabla}u \;+\; \left(
\begin{array}{c} 0 \\ 0 \\ 2\partial_1 \alpha
\end{array} \right)
\end{eqnarray}
which indeed satisfies
\begin{equation}
\vec{\nabla} h\;+\;\vec{\nabla} \times \vec{f}\;\equiv\;0
\end{equation}
(cp. (3)), due to $\alpha (x_1,x_2)$ being the real part of a
holomorphic function (i.e. satisfying $\partial_{11}\alpha +
\partial_{22}\alpha = 0)$; this means that (37)/(33)$_{m=1}$ describes
a hypersurface motion, (39), whose time-function satisfies (cp. (12))
\begin{equation}
\vec{\nabla} \left( ( c- 2\partial_2\alpha)^2 \ \vec{\nabla} t
  (\vec{x})\right) \;=\; 0
\end{equation}
- -- corresponding to a time-harmonic hypersurface motion in a
\ub{curved}, only conformally flat, space with metric
\begin{equation}
\tilde{\zeta}_{ij} (x)\;=\; (c-2\partial_2\alpha)^4 \ \delta_{ij}\ .
\end{equation}

\noindent {\bf Acknowledgement :} I am grateful to 
M. Bordemann for valuable discussions.

\vspace{2cm}

\noindent {\bf References}
\begin{itemize}
\item[[1]] J. Hoppe; {\sl Higher Dimensional Integrable Systems from
    Multilinear Evolution Equations}; ETH-TH/96-04.
\item[[2]] M. Bordemann, J. Hoppe; {\sl Diffeomorphism Invariant
    Integrable Field Theories and Hypersurface Motions in Riemannian
    Manifolds}; ETH-TH/95-31, FR-THEP-95-26.
\item[[3]] J. Hoppe; {\sl Lectures on Integrable Systems}, Springer
  Verlag 1992.
\item[[4]] N. Hitchin; Comm. Math. Phys. \ub{89} (1989) 145.
\end{itemize}

\end{document}